\documentclass[aps, prl, preprint, amsmath, amssymb]{revtex4}

\begin{document}

\title{\bf Gauge Fields and Unparticles}
\author{A. Lewis Licht}
\affiliation{Dept. of Physics\\U. of Illinois at Chicago\\Chicago, 
Illinois 60607}
\email{licht@uic.edu}

\begin{abstract}
    We show that a rigorous path integral method of introducing gauge fields in the
    UnParticle lagrangian leads to somewhat different and more 
    complicated vertexes than 
    those currently used.  
\end{abstract}
\maketitle
\section{Introduction}\label{S:intro}
The idea of a scalar field that represents a particle of indefinite mass, 
introduced by Georgi~\cite{Georgi-1}~\cite{Georgi-2}, was extended to 
a gauged field by Terning et al~\cite{Terning-1}.  The unparticle 
action was taken in the nonlocal form:

\begin{equation}
I = \int {d^4 xd^4 y\bar \psi \left( x \right)K\left( {x - y} \right)} \psi \left( y \right)
\end{equation}

Where K denotes the inverse of the Unparticle's propagator.  To 
include a gauge field A, an additional term is included:
\begin{equation}
U\left( {x,y,\gamma } \right) = P\left[ {\exp \left( { - ig\int_{x,\gamma }^y {A_\mu  \left( w \right)dw^\mu  } } \right)} \right]
\end{equation}

Here P denotes a path ordering, and $\gamma$ denotes a path from x to 
y.  Then
\begin{equation}\label{E:KUeq}
I = \int {d^4 xd^4 y\bar \psi \left( x \right)K\left( {x - y} \right)} U\left( {x,y,\gamma } \right)\psi \left( y \right)
\end{equation}

Terning et al~\cite{Terning-1} do not specify the path $\gamma$, but 
they make the assumption that it is always such that
\begin{equation}\label{E:partialU}
\frac{\partial }
{{\partial y^\mu  }}U\left( {x,y,\gamma } \right) =  - igU\left( {x,y,\gamma } \right)A_\mu  \left( y \right)
\end{equation}

This is a very old idea that goes back to 
Mandlestam~\cite{Mandlestam}, but it can not be quite correct.  It 
requires that for all x and y, the path that goes from x to y + dy must have gone first from x 
to y. However, the paths between all point pairs must be exactly 
defined before the integral in Eq.~(\ref{E:KUeq}) can be calculated.  
Whatever the definition of the path from x to y + dy, it cannot be 
expected to have gone through y just because someone wishes to 
compute the derivative.  

In the following, we investigate the consequences of defining the 
path integral between any two points x and y as the straight line 
from x to y.  We show that the resulting vertexes satisfy the 
Ward-Takahashi identities ~\cite{Ward-Takahashi}, but they do lead to 
vertexes that are rather more complicated than those found in 
Ref.~\cite{Terning-1}.  In a later work, we will show that a 
Terning-type vertex can be obtained by a different method of 
introducing gauge fields into the Unparticle action.

\section{Straight Line Path}\label{S:SLP}

Choosing the path as the straight line from x to y leads to
\begin{equation}
U\left( {x,y} \right) = P\left[ {\exp \left( { - ig\int_0^1 {A_\mu  \left( {w\left( \lambda  \right)} \right)dw^\mu  \left( \lambda  \right)} } \right)} \right]
\end{equation}
where
\begin{equation}
w^\mu  \left( \lambda  \right) = \left( {1 - \lambda } \right)x^\mu   + \lambda y^\mu  
\end{equation}
The UP-gauge-UP vertex is defined by
\begin{equation}
ig\Gamma ^\mu  \left( {y,x,z} \right) =  - \left. {\frac{{\delta ^3 I}}
{{\delta A_\mu  \left( x \right)\delta \psi \left( y \right)\delta \bar \psi \left( z \right)}}} \right|_{A = 0} 
\end{equation}
Fourier transforming:
\begin{equation}
ig\Gamma ^\mu  \left( {p,q,p + q} \right)\left( {2\pi } \right)^4 \delta \left( {p' - p - q} \right) = \int {d^4 xd^4 yd^4 z\operatorname{e} ^{i\left( {p'z - py - qx} \right)} ig\Gamma ^\mu  \left( {y,x,z} \right)} 
\end{equation}
Using
\begin{equation}
\left. {\frac{\delta }
{{\delta A_\mu  \left( w \right)}}U\left( {x,y} \right)} \right|_{A = 0}  =  - ig\int\limits_0^1 {d\lambda \delta \left( {w - \left( {1 - \lambda } \right)x - \lambda y} \right)\left( {y^\mu   - x^\mu  } \right)} 
\end{equation}
and with $S \left( k \right)$ denoting the Unparticle propagator in 
momentum space,
\begin{equation}
K\left( {x - y} \right) = \int {\frac{{d^4 k}}
{{\left( {2\pi } \right)^4 }}S^{ - 1} \left( k \right)} e^{ik\left( {x - y} \right)} 
\end{equation}
we get
\begin{equation}
\Gamma ^\mu  \left( {p,q,p + q} \right) =  - i\int\limits_0^1 {d\lambda \left. {\frac{\partial }
{{\partial k_\mu  }}S^{ - 1} \left( k \right)} \right|} _{k =  - \left( {p + \lambda q} \right)} 
\end{equation}
We show that this satisfies the Ward-Takahashi identity.  We consider 
first the scalar UP case, where the propagator depends on k through $s 
= k^2  = p^2  + 2\left( {p \cdot q} \right)\lambda  + q^2 \lambda ^2 
$. Then
\begin{equation}
\Gamma ^\mu   = 2i\int\limits_0^1 {d\lambda \left( {p^\mu   + \lambda q^\mu  } \right)} \frac{{dS^{ - 1} }}
{{ds}}
\end{equation}
and with$\frac{{ds}}
{{d\lambda }} = 2\left( {p \cdot q + \lambda q^2 } \right)$
 we get
 \begin{equation}
q^\mu  \Gamma _\mu   = i\int\limits_0^1 {d\lambda \frac{{ds}}
{{d\lambda }}\frac{{dS^{ - 1} }}
{{ds}}}  = i\left[ {S^{ - 1} \left( {p + q} \right) - S^{ - 1} \left( p \right)} \right]
\end{equation}
the WT relation.  If now the UP is a fermion, then
\begin{equation}
S^{ - 1}  = \gamma ^\mu  k_\mu  g\left( s \right)
\end{equation}
and
\begin{equation}
\Gamma ^\mu   =  - i\int\limits_0^1 {d\lambda \left[ {\gamma ^\mu  g + 2\gamma ^\alpha  \left( {p_\alpha   + \lambda q_\alpha  } \right)\left( {p^\mu   + \lambda q^\mu  } \right)\frac{{dg}}
{{ds}}} \right]} 
\end{equation}
then
\[
\begin{gathered}
  q^\mu  \Gamma _\mu   =  - i\int\limits_0^1 {d\lambda \gamma ^\alpha  \left[ {q_\alpha  g + \left( {p_\alpha   + \lambda q_\alpha  } \right)\frac{{dg}}
{{d\lambda }}} \right]}  \\ 
   = i\gamma ^\alpha  \left[ {\left( {p_\alpha   + q_\alpha  } \right)g\left( {\left( {p + q} \right)^2 } \right) - p_\alpha  g\left( {p^2 } \right)} \right] \\ 
   = i\left[ {S^{ - 1} \left( {p + q} \right) - S^{ - 1} \left( p \right)} \right] \\ 
\end{gathered} 
\]
the WT identity.

\section{The Vertex Integral}\label{S:TVI}

It is generally assumed ~\cite{Georgi-1} ~\cite{Georgi-2} that the Fourier transform of 
the inverse propagator goes as a power of the invariant momentum squared:
\begin{equation}
S^{ - 1} (k) \doteq \left( {k^2 } \right)^\nu  
\end{equation}
where $\nu = 2 - d_{u}$ , $d_{u}$ being the unparticle dimension. We 
therefore need to find the integral
\begin{equation}
f_\nu  \left( {p,p'} \right) = \int_0^1 {s^\nu  d\lambda } 
\end{equation}

This can be written as 

\begin{equation}\label{E:fnu}
f_\nu   = \frac{1}
{{2\sqrt {q^2 } }}\int_{s_0 }^{s_1 } {ds\frac{{s^\nu  }}
{{\sqrt {s - A} }}} 
\end{equation}

where $s_0  = p^2 $ , $s_1  = p'^2 $ and

\begin{equation}
A = p^2  - \frac{{\left( {p \cdot q} \right)^2 }}
{{q^2 }}
\end{equation}

If $ \nu $ is not a half integer, the integral in Eq. 
(~\ref{E:fnu}) can be done as an infinite series, giving

\begin{equation}
f_\nu  \left( {p,p'} \right) = g_\nu  \left( {p'^2 ,A} \right) - g_\nu  \left( {p^2 ,A} \right)
\end{equation}

where
\begin{equation}
g_\nu  \left( {s,A} \right) = \frac{{s^{\nu  + \frac{1}
{2}} }}
{{2\sqrt {q^2 } }}\sum\limits_{k = 0}^\infty  {\frac{{\left( {2k - 1} \right)!!}}
{{2^k k!\left( {\nu  - k + \frac{1}
{2}} \right)}}} \left( {\frac{A}
{s}} \right)^k 
\end{equation}

\section{The Scalar Vertex}\label{S:TSV}

In this section we show another way of calculating the vertex 
integral and derive the vertex for a scalar unparticle. The vertex is 
given in terms of the vertex integral as

\begin{equation}
\Gamma ^\mu   = 2i\left. {\frac{\partial }
{{\partial p^\mu  }}f_\nu  } \right|_q 
\end{equation}

where the scalar integral is 
\begin{equation}
f_\nu   = \int_0^1 {d\lambda \left( {s\left( \lambda  \right)} \right)^\nu  } 
\end{equation}

with $s = \left( {p + \lambda q} \right)^2 $ , this can be expanded 
in a Taylor series in $ \lambda $ and then integrated to give
\begin{equation}
f_\nu   = A^\nu  \sum\limits_{k = 0}^\infty  {\frac{{\Gamma \left( {\nu  + 1} \right)}}
{{k!\left( {2k + 1} \right)\Gamma \left( {\nu  - k + 1} \right)}}} \left( {\frac{{q^2 }}
{A}} \right)^k \left[ {\left( {1 + B} \right)^{2k + 1}  - B^{2k + 1} } \right]
\end{equation}
where
\[
\begin{gathered}
  A = p^2  - \frac{{\left( {p \cdot q} \right)^2 }}
{{q^2 }} \\ 
   = \frac{{p^2 p'^2  - \left( {p \cdot p'} \right)^2 }}
{{\left( {p' - p} \right)^2 }} \\ 
\end{gathered} 
\]
and
\[
\begin{gathered}
  1 + B = \frac{{p' \cdot q}}
{{q^2 }} \\ 
  B = \frac{{p \cdot q}}
{{q^2 }} \\ 
\end{gathered} 
\]
This can be expressed in terms of the hypergeometric functions

\begin{equation}
{}_2F_1 \left( {\alpha ,\beta ,\gamma ;z} \right) = 1 + \frac{{\alpha \beta }}
{\gamma }z + \frac{{\alpha \left( {\alpha  + 1} \right)\beta \left( {\beta  + 1} \right)}}
{{\gamma \left( {\gamma  + 1} \right)2!}}z^2  + ....
\end{equation}

as

\begin{equation}
f_\nu   = A^\nu  \left[ {\left( {1 + B} \right){}_2F_1 \left( {\frac{1}
{2}, - \nu ,\frac{3}
{2}; - \frac{{q^2 }}
{A}\left( {1 + B} \right)^2 } \right) - B{}_2F_1 \left( {\frac{1}
{2}, - \nu ,\frac{3}
{2}; - \frac{{q^2 }}
{A}B^2 } \right)} \right]
\end{equation}

Defining
\begin{equation}
Q_\nu  (z) = {}_2F_1 \left( {\frac{1}
{2}, - \nu ,\frac{3}
{2}; - \frac{{q^2 }}
{A}z^2 } \right)
\end{equation}

and using
\begin{equation}
\left. {\frac{{\partial B}}
{{\partial p_\mu  }}} \right|_q  = \frac{{q^\mu  }}
{{q^2 }}
\end{equation}
\[
\begin{gathered}
  \left. {\frac{{\partial A}}
{{\partial p_\mu  }}} \right|_q  = 2\left( {p^\mu   - \frac{{\left( {p \cdot q} \right)}}
{{q^2 }}q^\mu  } \right) \\ 
   = \frac{2}
{{q^2 }}\left[ {p^\mu  \left( {p' \cdot q} \right) - p'^\mu  \left( {p \cdot q} \right)} \right] \\ 
\end{gathered} 
\]
\begin{equation}
C_\nu   = \left( {1 + B} \right)Q_\nu  \left( {1 + B} \right) - BQ_\nu  \left( B \right)
\end{equation}
we get
\begin{equation}
\Gamma ^\mu   = 2i\left\{ {\frac{{q^\mu  }}
{{q^2 }}\left[ {p'^{2\nu }  - p^{2\nu } } \right] + \frac{{2\nu A^{\nu  - 1} }}
{{q^2 }}\left[ {p^\mu  \left( {p' \cdot q} \right) - p'^\mu  \left( 
{p \cdot q} \right)} \right]C_{\nu - 1}  } \right\}
\end{equation}

This is considerably more complicated than the result found in Ref.~\cite{Terning-1}.

When $ \nu = 1 $, this reduces to
\[
\Gamma ^\mu   = 2i\left( {p'^\mu   + p^\mu  } \right)
\]

The expected result for a scalar particle.

\section{Conclusions}\label{S:Conc}

We find that a rigorous application of the path integral method of 
introducing gauge fields into the unparticle Lagrangian leads to 
vertexes that are connsiderably more complicated than those found by 
Terning et al.~\cite{Terning-1} We will show in a later work that there 
is another method of combining gauge fields and unparticles does lead 
to the Terning result.

\section{Acknowledgements}
I wish to express my thanks to Wai-Yee Keung for 
interesting me in this problem.

\end{document}